\documentclass[showpacs,aps,prb,twocolumn]{revtex4}
\usepackage{graphicx}
\usepackage{amsmath}
\begin{document}
\bibliographystyle{apsrev}
\title{Macroturbulent Instability of the Flux Line Lattice in
Anisotropic Superconductors}
\author{L. M. Fisher}
\affiliation{All-Russian Electrical Engineering Institute, 12
Krasnokazarmennaya Street, 111250 Moscow, Russian Federation}
\author{T. H. Johansen}
\affiliation{Department of Physics, University of Oslo, P.O. Box
1048, Blindern, 0316 Oslo 3, Norway}
\author{A. L. Rakhmanov}
\affiliation{Institute for Theoretical and Applied Electrodynamics
Russian Academy of Science, 13/19 Izhorskaya Street, 127412
Moscow, Russian Federation}
\author{A. A.~Levchenko, V. A.~Yampol'skii}
\affiliation{A.~Ya.~Usikov Institute for Radiophysics and
Electronics Ukrainian Academy of Science,12 Proskura Street, 61085
Kharkov, Ukraine}


\begin{abstract}
A theory of the macroturbulent instability in the system
containing vortices of opposite directions (vortices and
antivortices) in hard superconductors is proposed. The origin of
the instability is connected with the anisotropy of the current
capability in the sample plane. The anisotropy results in the
appearance of tangential discontinuity of the hydrodynamic
velocity of vortex and antivortex motion near the front of
magnetization reversal. As is known from the classical
hydrodynamics of viscous fluids, this leads to the turbulization
of flow. The examination is performed on the basis of the
anisotropic power-law current-voltage characteristics. The
dispersion equation for the dependence of the instability
increment on the wave number of perturbation is obtained, solved,
and analyzed analytically and numerically. It is shown that the
instability can be observed even at relatively weak anisotropy.
\end{abstract}
 {74.25.Ha, 74.60.Ge,
74.60.Jg} 
 \maketitle

\section{Introduction}

Magnetic flux dynamics in type-II superconductors has been
extensively studied since the end of the 50-s, starting with the
pioneering work by A.~A.~Abrikosov. Primary attention was given to
hard superconductors, whose magnetic characteristics are defined
by the presence of vortex pinning centers. The main features
characterizing the nonuniform penetration of magnetic flux into
such systems were revealed and studied; various theoretical models
of electrodynamic processes in superconductors were suggested. An
avalanche of new researches in this field was triggered by the
discovery of high-$T_c$ superconductivity (HTS), which brought
into play thermal fluctuations and the strong anisotropy of
superconductors. Different types of the phase transitions
(melting, glass state transitions, etc.\ ) in the flux line
lattice (FLL) have been discovered, studied, and explained. Many
of the newly obtained results are described in comprehensive
review papers~\cite{blat,2}.

The use of high-resolution magneto-optical (MO) technique enabled
an in-depth study of the dynamics of magnetic flux in
superconductors. One of the most important features revealed by
means of this method is the fractal structures of thermally
activated flow of the magnetic flux. Such fractal structures arise
usually due to the development of the characteristic instabilities
such as macroturbulence in 1-2-3 systems~\cite{vl,ind,joh} and
dendrite instability in magnesium deboride~\cite{dendrite}.
However, these dynamic instabilities in FLL have not been so far
properly investigated.

Perhaps, the macroturbulence is the most dramatic phenomena
observed in the dynamics of the magnetic flux in HTS. It appears
like a turbulization of the FLL motion in a sample near the front
of magnetization reversal that separates regions of the vortices
of opposite directions (vortices and antivortices). Note that the
macroturbulence was only revealed in single-crystal samples of the
1--2--3 system. Its essence is as follows. When a magnetic flux is
trapped in a superconductor and a moderate field of a reverse
direction is subsequently applied, a boundary of zero flux density
will separate the regions containing vortices and antivortices.
For definiteness, we apply the term ``antivortices'' to the
vortices whose direction coincides with that of the external
magnetic field, and the term ``vortices'', to the vortices which
were originally present in the sample, prior to switching on the
magnetic field of negative sign. At some range of magnetic fields
and temperatures, this flux-antiflux distribution becomes
unstable. A disordered motion of magnetic flux arises at the front
of magnetization reversal, which resembles a turbulent fluid flow.
This process rapidly develops in time and is accompanied by the
emergence of channels via which the antivortices penetrate into
the region occupied by the vortices. In other words, the front of
magnetization reversal takes on a finger-like shape. The
annihilation of vortices and antivortices occurs at the front, and
the process of macroturbulence is soon terminated after a complete
disappearance of the vortices. This pattern of penetration of the
magnetic flux differs qualitatively from the steady-state slow
motion of the front of magnetization upon initial switching on of
the magnetic field, when vortices of only one direction are
present in the sample. Note that the characteristic times of
instability development amount to several seconds and more, and
emerging spatial structures are macroscopic, i.\ e., they contain
a large number of single vortices.

The described instability cannot be understood within the
framework of the critical state model or conventional models for
the thermoactivated flux relaxation. At the same time, this
phenomenon is a close analogue to the turbulence in the usual
hydrodynamics. So, the study of the phenomenon is of general
physical interest. Moreover, this investigation has evident
application aspects, since the macroturbulence can affect the ac
losses, magnetic noise, and relaxation phenomena in
superconducting devices.

An attempt to explain this remarkable behavior of the
flux-antiflux interface was undertaken in paper by Bass et al.
~\cite{bass}, where the instability was attributed to a thermal
wave generated by the heat release due to the vortex-antivortex
annihilation. Unfortunately, this mechanism can hardly be
relevant. Indeed, the annihilation energy is small and a
corresponding temperature rise is negligible~\cite{prl,jetp}.

Another physical pattern for the emergence of macroturbulence was
discussed by Vlasko-Vlasov et al.~\cite{hole}. They draw attention
to the fact that the process of annihilation of a
vortex-antivortex pair may be accompanied by the formation of
spatial domains free of vortices (the so-called Meissner holes).
It was assumed that the presence of such domains may cause
instability due to high currents which have to flow in the
vicinity of such a Meissner hole. Yet, the authors of
Ref.~\onlinecite{hole} did not describe the possible physical
instability mechanism.

An explanation of the macroturbulence should focus on the
experimental fact that the instability was reported for
$\mathrm{YBa_2Cu_3O_{7-\delta}}$ and other 1--2--3 single crystals
only, which are characterized by the anisotropy in the basal
\textbf{ab} plane. This anisotropy can be related to a specific
crystallographic structure of these HTS and, in particular, to
twin boundaries~\cite{TB}. The electromagnetic instabilities of
the critical and resistive states in anisotropic superconductors
were considered by Gurevich~\cite{G1,G2}. However, these results
cannot be directly applied to the explanation of the turbulent
instability.

An alternative approach to understanding the mechanism of the
macroturbulence was elaborated in Refs.~\onlinecite{prl,jetp}
taking into account of the specific features of the flux motion in
the anisotropic superconductors. The anisotropy gives rise to the
motion of the flux lines at some angle with respect to the Lorentz
force direction. For example, in the presence of twin boundaries,
vortices and antivortices move preferably along these guiding
boundaries (the so-called guiding effect~\cite{guid1,guid2}). In
general, the angle between the twins and the crystal grains is
around $45^\circ$. As a result, the flux lines move at some angle
with respect to the magnetization reversal front. In our opinion,
it is just this circumstance that is of paramount importance to
ascertain the nature of macroturbulence. The vortices and
antivortices are forced to move towards each other in such a way
that the tangential component of their velocity becomes
discontinuous at the flux-antiflux interface. According to a
classical paper of Helmholtz, a stationary hydrodynamical flow can
be unstable and turbulent under such conditions. It was
shown~\cite{prl} that a purely hydrodynamic approach which takes
into account the anisotropic viscosity of the flux flow can
provide the basis to gain an appropriate insight into the nature
of the macroturbulence. In particular, the macroturbulence is
usually observed within a rather narrow temperature window in the
vicinity of 40--50~K~\cite{bazil}. As was demonstrated in
Ref.~\onlinecite{prl}, the temperature window is considerably
wider for heavily twinned samples in which a more pronounced
anisotropy can be expected.

The model developed in Refs.~\onlinecite{prl,jetp} is based on the
analysis of the viscous magnetic flux flow in anisotropic
superconductors. In the framework of this model the dependence of
the viscosity coefficient of the flux line lattice on the flux
velocity was neglected. This approximation corresponds  evidently
to the description of the superconductor in terms of the linear
current-voltage ($I-V$) characteristics. This idealized model
allows one to describe qualitatively the effect in question.
However, it predicts that the macroturbulence is observable for
unrealistically high values of the anisotropy parameter. Note that
the regime of the viscous magnetic flux flow is realized at
extremely high current densities that cannot be achieved under
experimental conditions used in the MO measurement. This is a
reason for using a more general model for the flux flow in the
present paper. Specifically, we take into account the dependence
of the viscosity coefficient on the flux flow rate. In other
words, we describe the electric properties of a superconductor by
a more realistic non-linear $I-V$ characteristics. We have
theoretically studied the instability of the magnetization
reversal front under the assumption that the current-voltage
characteristic is a power-law function with an exponent $m\geq 1$.
We show that, even at a comparatively weak anisotropy of the
current-voltage characteristics, the flow of a system of vortices
and antivortices in a superconductor can be unstable. A
preliminary results of the present study was published in the
letter~\cite{pisma}.

\section{Main equations}

\begin{figure}[!ht]
\includegraphics[width=0.4\textwidth,height=0.4\textwidth]{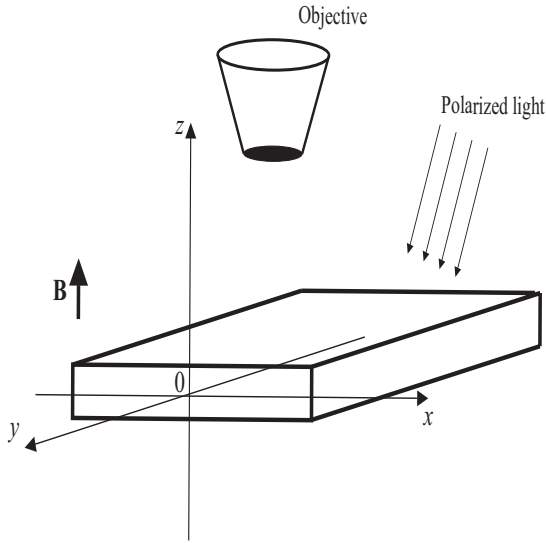}
\caption{\label{f1} Geometry of the problem.}
\end{figure}

The macroturbulence is usually observed in plate-like single
crystals placed in a transverse magnetic field. The MO image
provides information about the distribution of the normal
component of magnetic induction. The penetration of an
electromagnetic field into a superconductor in such a geometry is
of interest \textit{per se} and was studied by numerous
researchers (see, for example, Refs.~\cite{G3,brandt}). However,
the turbulent behavior of magnetic flux is not a geometric effect.
It was observed in thin plates and in single crystals with a low
demagnetizing factor as well. Frello et al.~\cite{joh} reported
the MO visualization of developed macroturbulence of vortex matter
in an Nd-123 crystal $3.1\times 2.5 \times 1.3$~mm$^3$ in size. In
the magnetic field, this sample was divided into three
magnetically unbound regions (each having smaller transverse
dimensions) in which the turbulence was developing independently.
Moreover, it should also be kept in mind that the instability is
often observed under conditions of the full penetration of
magnetic flux into the sample when the difference in the
distribution of induction in the cases of longitudinal and
transverse geometry turns out to be not too
significant~\cite{brandt}. Therefore, for a qualitative
description we can restrict ourselves by the simplest longitudinal
geometry. So, consider an infinite superconducting plate of
thickness $2d$ in the external magnetic field ${\vec H}$ directed
parallel to the sample surface along the $z$-axis. The $x$-axis is
perpendicular to the plate and the $x$-axis origin, $x=0$, is
placed in the plate center (see Fig.~\ref{f1}).
Let the external magnetic field $H$ first increase
up to some value much higher than the lower critical magnetic
field $H_{c1}$. Then $H$ is lowered through zero to a negative
constant value whose modulus is likewise higher than $H_{c1}$.
Under such conditions, two kinds of vortices exist inside the
sample. The first one with the magnetic flux directed along the
positive $z$-axis has penetrated into the sample under the
magnetic field rise ({\it vortices\/}). The vortices occupy the
central part of the superconductor. The second group of vortices
entering the sample after the external field has changed its sign
is characterized by the flux directed oppositely ({\it
antivortices\/}). The antivortices are located in two peripheral
regions arranged symmetrically relative to the median plane $x=0$
of the plate.

In real crystals, the vortices and antivortices are pinned by
defects. The vortex motion occurs due to the action of the Lorentz
force and (at non-zero temperature $T$) due to the thermoactivated
flux creep. In the framework of the macroscopic description of the
flux flow we can take this effect into consideration by making use
of the non-linear $I-V$ characteristics~\cite{blat}. There exists
an additional reason for the flux line motion in the situation
under study. The vortices and antivortices can annihilate near the
plane of zero magnetic induction. This phenomenon gives rise to
diminishing the vortices number in the bulk and to entering new
antivortices through the sample surface. As a result, the line of
zero induction moves towards the sample center with time. We will
use the phenomenological (hydrodynamic-like) equations to describe
the flux line motion. This approach is valid if all spatial scales
of the problem are much larger than the FLL constant $d_f$. In
particular, the characteristic spatial scale of the
macroturbulence $l_c$ should be large, $l_c\gg d_f$.

\begin{figure}[!h]
\includegraphics[width=0.3\textwidth,height=0.3\textwidth]{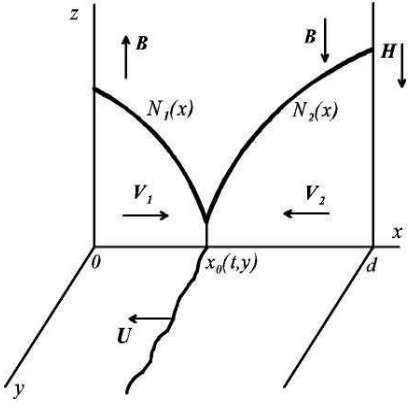}
\caption{\label{f2} Flux distribution in one half of an infinite
slab $(|x|\leq d)$ containing trapped vortices of density $N_1(x)$
in the central region $|x| \leq x_0$, and antivortices of density
$N_2(x)$ penetrating from the outside. The other symbols are
defined in the text.}
\end{figure}

Let us denote the density of the vortices and antivortices as
$N_1(x,y)$ and $N_2(x,y)$, respectively. The relation between
vortex densities $N_{\alpha}(x,y)$ and the magnetic induction
$B(x,y)$ is evident,
\begin{equation} \label{1}
N_{\alpha}(x,y) = s_{\alpha}B(x,y)/\Phi_0, \qquad \alpha = 1, 2
\end{equation}
where $s_1=1$, $s_2=-1$ and $\Phi_0$ is the magnetic flux quantum.
>From the symmetry of the problem, it is sufficient to consider
only the region $0<x<d$. Fig.~\ref{f2} shows schematically the
spatial distributions of $N_1(x)$ and $N_2(x)$. The vortex and
antivortex densities should evidently satisfy the continuity
equation,
\begin{equation}  \label{2}
\frac{\partial N_\alpha}{\partial t} + \mathrm{div}(N_\alpha{\vec
V}_\alpha)=0,
\end{equation}
where ${\vec V}_\alpha$ denotes the hydrodynamic velocities of the
vortices and antivortices.

In the hydrodynamic approximation, there are two equivalent
approaches to find the vortex and antivortex velocities. The usual
dynamic approach, where ${\vec V}_\alpha$ is defined by the
viscosity equation for FLL, which takes account of the Lorentz
force, was applied in Refs.~\onlinecite{prl,jetp}. The alternative
approach implies the use of the $I-V$ characteristics,
\begin{equation}\label{I-V}
{\vec J}={\vec J(\vec E)},
\end{equation}
where ${\vec J}$ and ${\vec E}$ are the macroscopic current
density and the electric field. We operate with the latter method
in this paper.

The velocities and electric field are interrelated by the usual
equation,
\begin{equation}  \label{3} {\vec E}=\frac{1}{c}{\vec
V}\times {\vec B}.
\end{equation}
Using Eq.~(\ref{1}), we can rewrite Eq.~(\ref{3}) in the form,
\begin{equation}  \label{4}
E_x=-\frac{N_{\alpha}s_{\alpha}\Phi_0}{c}V_{\alpha y}, \quad
E_y=\frac{N_{\alpha}s_{\alpha}\Phi_0}{c}V_{\alpha x}.
\end{equation}
In principle, we can use the $I-V$ characteristics in general form
(\ref{I-V}). However, in so doing the obtained results are rather
cumbersome and their physical understanding is not evident. To
avoid this inconvenience, we use here some particular power-law
$I-V$ characteristics for anisotropic superconductors. Let the
main anisotropy axes be $X$ and $Y$ (e. g., the $X$-axis is
directed along twins, while the $Y$-axis is across them). We
suppose that the components of the current density vector along
these directions are the odd functions of the electric field and
for positive $E$ can be presented as
\begin{equation}\label{5}
J_X=\frac{1}{\varepsilon}J_c\left(\frac{E_X}{E_0} \right)^{1/m},
\qquad  J_Y=J_c\left(\frac{E_Y}{E_0} \right)^{1/m},
\end{equation}
where $J_c$ is the critical current density along the $Y$ axis,
$m>1$ is the corresponding exponent, $\varepsilon<1$ is the
anisotropy parameter for the critical current components and the
value of $J_c$ is defined as the current density $J_Y$ at $E=E_0$
(usually, $E_0$ is accepted as 1~$\mu$V/cm). At $m=1$ the $I-V$
characteristics (\ref{5}) correspond to the viscous flux flow
regime used for the macroturbulence analysis in
Ref.~\onlinecite{prl}. In the limiting case $m\gg 1$, these
characteristics are in accordance with the critical state model
proposed by Clem~\cite{clem} for the description of the
electrodynamics of anisotropic superconductors.

As was mentioned above, the anisotropy in \textbf{ab} plane arises
in 1--2--3 single crystals mainly due to the existence of twin
boundaries. In a real experimental situation, these boundaries
make angles of about $45^{\circ}$ with the axes $x$ and $y$.
Therefore, we assume the angle between two coordinate systems,
\textbf{xy} and \textbf{XY}, to be equal to $45^{\circ}$. Now,
using Eqs.~(\ref{4}), (\ref{5}) and the Maxwell equation
$\nabla\times\vec{B}=4\pi\vec{J}/c$ we can derive the equations
interrelating the vortex density and velocity components,
\begin{equation}\label{6}
\frac{\partial N_{\alpha}}{\partial x}-\frac{\partial
N_{\alpha}}{\partial y}=\frac{4\pi\sqrt{2}J_c}{c\Phi_0
\varepsilon}\left[\frac{N_\alpha\Phi_0}
{cE_0\sqrt{2}}\left(-V_{x\alpha}+V_{y\alpha}\right)
\right]^{^1/m},
\end{equation}
\vspace*{-4mm}
\[
\frac{\partial N_{\alpha}}{\partial x}+\frac{\partial
N_{\alpha}}{\partial
y}=-\frac{4\pi\sqrt{2}J_c}{c\Phi_0}\left[\frac{N_\alpha\Phi_0}
{cE_0\sqrt{2}}\left(V_{x\alpha}+V_{y\alpha}\right) \right]^{^1/m}.
\]

To solve the problem, we must formulate the boundary conditions at
the sample surface, as well as at the interface between the
domains occupied by the vortices and antivortices. We will first
discuss the conditions at the sample boundaries. Ignoring the
induction jump on the surface (this may be done in the case of
fairly high values of $H$, $H\gg H_{c1}$), one can derive
\begin{equation}\label{bc}
N_2(d)=N_2(-d)=H/\Phi_0.
\end{equation}
Since only the right-hand part $0< x < d$ of the sample is
treated, we will replace the condition $N_2(-d) = H/\Phi_0$ by the
requirement,
\begin{equation}\label{5a}
 V_\alpha(0)=0,
\end{equation}
that immediately follows from the symmetry of the problem.

Now we turn to the boundary conditions at the moving
vortex-antivortex interface. In general case, the position of the
interface depends on time $t$ and the $y$ coordinate. The equation
for this surface can be presented as $x=x_0(y,t)$. Then, one can
define the interface velocity $\vec{U}$ as a vector normal to this
surface,
\[
U_x=\frac{\partial x_0}{\partial t}\frac{1}{1+(\partial
x_0/\partial y)^2},
\]
\begin{equation}\label{interU}
U_y=-\frac{\partial x_0}{\partial t}\frac{\partial x_0/\partial
y}{1+(\partial x_0/\partial y)^2}.
\end{equation}
In general, the velocity $\vec{U}$, just as $x_0(y,t)$, depends on
time and the coordinate $y$, $\vec{U}=\vec{U}(y,t)$.

In the coordinate system which moves with the interface, the total
flux of the vortex and antivortex densities through the interface
should vanish due to the evident vortex conservation law. This
means that the following boundary condition should be fulfilled at
$x=x_0$,
\begin{equation}\label{b1}
N_1 \left(\vec{V_1}-\vec{U} \right)_n + N_2
\left(\vec{V_2}-\vec{U} \right)_n=0,
\end{equation}
where the subscript $n$ denotes the vector components transverse
to the interface. The components of the corresponding normal unit
vector $\vec{\nu}$ are
\[
\nu_x=\frac{1}{\sqrt{1+\left(\partial x_0/\partial y\right)^{2}}},
\]
\begin{equation}\label{normal}
\nu_y=-\frac{\partial x_0/\partial y}{\sqrt{1+\left(\partial
x_0/\partial y\right)^{2}}}.
\end{equation}

The second boundary condition should define the rate of the
vortex-antivortex annihilation. It is obvious that this rate goes
to zero if the density of vortices or antivortices at the
interface between them is zero. Then, following the conventional
approach to describing such kinetic processes, we represent the
rate of annihilation to be proportional to the product of the
vortex and antivortex densities,
\begin{equation}  \label{b2}
N_1({\vec V}_1 -{\vec U})_n = RN_1N_2.
\end{equation}
A similar model for the annihilation process was used in
Ref.~\onlinecite{bass}. The finite size region where the
annihilation occurs was supposed to exist in the sample. The
annihilation rate was assumed to be proportional not only to the
product $N_1$ and $N_2$ but also to the relative velocity
$|V_1-V_2|$ of the vortices and antivortices. Contrary to
Ref.~\onlinecite{bass}, we believe that the region where vortices
and antivortices coexist and annihilate is small enough and is
much less as compared not only to the sample sizes but also to the
characteristic spatial scale $l_c$ of the macroturbulence. This
assumption can be confirmed by MO images of the macroturbulent
instability~\cite{vl,ind,joh}. It means, in particular, that the
annihilation constant $R$ is not too small. Thus, the
flux-antiflux boundary can be represented by the surface
$x=x_0(y,t)$.

Finally, we assume that the average magnetic induction in the
neighborhood of the interface is zero, i.e.,
\begin{equation}  \label{b3}
N_1=N_2
\end{equation}
at $x=x_0 (y,t)$. One can readily demonstrate that this condition
directly follows from Eq.~(\ref{2}) and relation (\ref{b1}), if it
was valid at the moment of emergence of antivortices into the
sample. In our case, $N_1 = N_2 = 0$ at the time moment when the
decreasing external magnetic field assumed the value of $H = 0$.

\section{Unperturbed profile of the vortex distribution}

The formulated set of equations and boundary conditions allow us,
in principle, to analyze the behavior of the vortex-antivortex
system. We assume that the unperturbed flux-antiflux boundary is a
straight line. A scheme of solving the problem consists in finding
the unperturbed vortex distribution $N_\alpha(x,t)$ and velocity
$V_\alpha(x,t)$. Using these base distributions, one should solve
a linearized set of equations for fluctuations $\delta
N_\alpha(x,y,t)$ and $\delta V_\alpha(x,y,t)$. This section is
devoted to the analysis of the base profile. Below we take into
account the dependence of the critical current density $J_c$ on
the magnetic induction $\Phi_0 N_\alpha$. For definiteness sake,
we choose it in the simplest form,
\begin{equation}\label{mfd}
 J_c=A/N_\alpha.
\end{equation}

To find the base profile of the vortex density it is necessary to
solve a complex set of partial differential
equations~(\ref{2})--(\ref{6}) along with the boundary
conditions~(\ref{bc})--(\ref{b3}) some of which are written at the
moving interface. A peculiar complication consists in the fact
that the velocity $U(t)$ of the motion of the vortex-antivortex
interface is unknown and should be found self-consistently.
Unfortunately, the problem does not have simple (with
$U=\mathrm{const} \ne 0$) automodel solutions because the
interface moves non-uniformly under the conditions being studied.

As the first approximation, let us calculate a stationary base
profile of another system where the velocity $U(t)$ equals zero. A
superconducting slab carrying the critical transport current is an
example where such a distribution of the vortices and antivortices
can be realized. Assuming $\partial N_{\alpha}/\partial t= 0$ in
Eq.~(\ref{2}), one can easily obtain the distributions $N_1(x)$
and $N_2(x)$,
\begin{eqnarray}  \label{stprof}
\displaystyle N_\alpha(x)=N_0\sqrt{1+s_\alpha C(x_0-x)/d},
 \\
\noindent
 C=\frac{8\pi \sqrt{2} d A}{c \Phi_0 N_0^2} \cdot
\left[\frac{\sqrt{2} \Phi_0 RN_0^2}{c E_0 (1+\varepsilon
^m)}\right]^{1/m},\nonumber
\end{eqnarray}
from Eqs.~(\ref{2})--(\ref{6}) and
conditions~(\ref{b1})--(\ref{b3}). Following the
experiments~\cite{vl,ind,joh} which show that the vortex density
$N_0$ at the interface is small with respect to its value
$N_2(d)=H/\Phi_0$ at the sample surface, we assume the constant
$C$ to be much higher than unity. Neglecting unity in the radicand
in Eq.~(\ref{stprof}) and using condition (\ref{bc}) one can
evaluate the vortex density $N_0$ at the interface,
\begin{equation}\label{N_0}
N_0 = N_\alpha (x=x_0) \sim \left(\frac{H}{H_p}\right)^m
\left(\frac{c E_0}{2^{(m+1)/2}\Phi_0 R}\right)^{1/2},
\end{equation}
\[
H_p=(8\pi d A \Phi_0 /c)^{1/2}.
\]

Expression~(\ref{stprof}) is the solution of the stationary
problem. We are interested in the base profile corresponding to
the moving interface. This motion leads to a distortion of the
profile with respect to Eq.~(\ref{stprof}). However, we assume the
velocity $U$ to be small with respect to the vortex velocity
$V_\alpha$, $U\ll V_\alpha$. As it can be readily shown, in this
case the base profile differs but slightly from the stationary
distribution~(\ref{stprof}).

It is convenient to perform the further analysis using the
dimensionless variables,
\begin{eqnarray}  \label{dv}
\displaystyle n_\alpha = N_\alpha/N_0, \quad \tau=t/t_0, \quad
t_0= \frac{\Phi_0^2 N_0^3}{8\pi A E_0},
\nonumber\\
\noindent \ \xi=x/L, \quad \zeta=y/L, \quad L=\frac{c \Phi_0
N_0^2}{4\pi \sqrt{2} A},
\nonumber \\
\noindent r=\frac{R N_0^2 \Phi_0}{\sqrt{2}c E_0}, \quad u=U t_0/L.
\end{eqnarray}
The normalization to the time-dependent value
$N_0=N_\alpha(x_0(t))$ is allowable here since we assume that the
instability develops much faster than noticeable changes in
$N_\alpha(x_0)$ and $U(t)$ occur.

Taking into account the smallness of the velocity $u(\tau)$ of the
flux-antiflux boundary, we can linearize the expression for the
base profile $n_\alpha(\xi,\tau)$ with respect to $u$. In so
doing, we get the equations for the first and second derivatives,
$n_\alpha ^{'}$ and $n_\alpha ^{''}$, of the unperturbed vortex
density near the interface,
\begin{eqnarray} \displaystyle \label{der1}
n_\alpha ^{'}(x=x_0)=-s_\alpha \rho \left(1+s_\alpha
\frac{u}{rm}\right),\\
\noindent \label{der2} n_\alpha ^{''}(x=x_0)=-(n_\alpha ^{'})^2 -
\frac{2u}{(1+\epsilon) m (n_\alpha ^{'})^{m-2}},\\
\noindent \epsilon=\varepsilon ^m, \quad \rho = \left(
\frac{2r}{1+\epsilon}\right)^{1/m}.\nonumber
\end{eqnarray}
These formula are used in the next section devoted to the analysis
of the instability.

\section{Instability in the vortex-antivortex system}

Let the dimensionless vortex density be a sum of the unperturbed
term $\tilde n_\alpha (\xi,\tau)$ and a fluctuation term,
\begin{equation}  \label{fl}
n_\alpha = \tilde n_\alpha + f_\alpha (\xi-\xi_0(\tau))\exp
(\lambda \tau +{\rm i} k \zeta ).
\end{equation}
The linearized boundary conditions should be written at the
perturbed interface,
\begin{equation}  \label{int}
\xi=\xi_0(\zeta, \tau)=\xi_0(\tau)+ \delta \xi \exp({\rm i}k\zeta
+\lambda \tau).
\end{equation}
It follows directly from Eq.~(\ref{b1}) that
\begin{equation}\label{delta}
\delta \xi = (f_1 - f_2)/2\rho.
\end{equation}

Substituting Eq.~(\ref{fl}) into Eq.~(\ref{2}) and neglecting the
terms proportional to $\epsilon ^2$, we derive the ordinary
differential equation with the coordinate-dependent coefficients
for the fluctuation $f_\alpha (\xi-\xi_0(\tau))$,
\[
f_\alpha ^{''} + 2f_\alpha ^{'} \left[\mathrm{i}k (1-2\epsilon)+
\frac{\tilde n_\alpha ^{'}}{\tilde n_\alpha} -  u
\frac{m-2}{m{\tilde n_\alpha ^{'(m-1)}}{\tilde n_\alpha
^{m}}}\right]-
\]
\[
\frac{2\lambda}{m {\tilde n_\alpha ^{'(m-1)}}{\tilde n_\alpha
^{m}}} - \frac{2u}{{\tilde n_\alpha^{'(m-2)}}{\tilde n_\alpha
^{(m+1)}}} -\frac{2u\mathrm{i}k(m-1)}{m{\tilde
n_\alpha^{'(m-1)}}{\tilde n_\alpha ^{m}}}-
\]
\begin{equation} \label{ode}
\frac{{\tilde n_\alpha^{' 2}}}{{\tilde n_\alpha ^{2}}}-
k^2+2\mathrm{i}k \frac{\tilde n_\alpha^{'}}{{\tilde n_\alpha }}=0.
\end{equation}
We assume the perturbation of the vortex density to be damped away
from the interface $\xi=\xi_0 (\tau)$ at short distances with
respect to the sample thickness $d$. This allows us to replace the
coordinate-dependent variables ${\tilde n_\alpha}(\xi)$ and
${\tilde n_\alpha}^{'} (\xi)$ in Eq.~(\ref{ode}) by their values
${\tilde n_\alpha}(\xi=\xi_0)$ and ${\tilde n_\alpha}^{'}
(\xi=\xi_0)$ at the interface (see Eq.~(\ref{der1})).

The solution of Eq.~(\ref{ode}) defines the exponential behavior
of the fluctuations $f_1$ and $f_2$,
\[
f_1(\xi - \xi_0)=f_1 \exp [p_1(\xi - \xi_0)],
\]
\begin{equation}\label{exp}
f_2(\xi - \xi_0)=f_2 \exp [p_2(\xi - \xi_0)],
\end{equation}
where
\begin{equation}\label{p12}
p_{1,2}= \pm \rho - \mathrm{i}k + \rho u/2r \pm \Omega_{1,2},
\end{equation}

\[
\Omega_{1,2}=\left[2\rho^2 + \frac{\mathrm{i}ku\rho}{mr}
+\frac{\lambda \rho}{mr}+ 4\epsilon k^2  \mp
\frac{\mathrm{i}ku\rho (m-1) u}{m^2 r^2} \pm \right.
\]
\begin{equation}\displaystyle \label{om12}
\left. \frac{2u\rho^2}{mr}  \mp \frac{\lambda\rho (m-1)u }{m^2
r^2}\right]^{1/2}, \qquad \mathrm{Re}\, \Omega_{1,2}>0.
\end{equation}
Here the upper and lower signs correspond to the coefficients
$p_1$ and  $p_2$, respectively.

Substituting Eqs.~(\ref{fl})--(\ref{om12}) into the boundary
conditions~(\ref{b1}), (\ref{b2}) gives two linear algebraic
homogeneous equations for $f_1$ and $f_2$,
\begin{widetext}
\begin{eqnarray} \displaystyle \label{he}
f_1 \left[-\frac{m}{2} {\tilde n_1}^{'(m-1)}(p_1 +
\mathrm{i}k(1-2\epsilon)+ {\tilde
n_1}^{'})-\frac{\mathrm{i}ku(1-3\epsilon)}{\rho}-u-\frac{\lambda}{\rho}
\right] +
\nonumber \\
f_2\left[-\frac{m}{2} {\tilde n_2}^{'(m-1)}(p_2 +
\mathrm{i}k(1-2\epsilon)+ {\tilde n_2}^{'})+
\frac{\mathrm{i}ku(1-3\epsilon)}{\rho}- u+\frac{\lambda}{\rho}
\right]=0,
\end{eqnarray}
\begin{eqnarray} \displaystyle \label{2he}
f_1 \left[-\frac{m}{2} {\tilde n_1}^{'(m-1)}(p_1 +
\mathrm{i}k(1-2\epsilon)+ {\tilde
n_1}^{'})+\frac{\mathrm{i}k(1-2\epsilon)}{4\rho} {\tilde
n_1}^{'m}-\frac{\lambda}{2\rho} -r -u\frac{m-1}{m}-\frac{u\epsilon
{\tilde n_1}^{'} }{2\rho}\right] +
\nonumber \\
 f_2 \left[\frac{\lambda}{2\rho}-
\frac{\mathrm{i}k(1-2\epsilon)}{4\rho} {\tilde n_1}^{'m}-  r
-\frac{u}{m}+\frac{u\epsilon {\tilde n_1}^{'} }{2\rho}\right]=0.
\end{eqnarray}
\end{widetext}
Here we took into account that the normal unit vector to the
perturbed interface is ${\vec \nu}=(1, \; -{\rm i} k\delta
\xi(\zeta,\tau))$.

Requiring that the determinant of set~(\ref{he}), (\ref{2he})
should vanish and omitting the terms of the order of $u^2\ll 1$,
one obtains the dispersion equation for the increment $\lambda$ at
different values of the wave number $k$,
\begin{equation}  \label{de}
\lambda=\frac{mr}{\rho}\left(\Omega^2-4\epsilon k^2 -\mathrm{
i}ku\frac{\rho}{mr} -2\rho^2\right).
\end{equation}
Here $\Omega$ is the root with $\mathrm{Re}\, \Omega >0$ of the
equation
\begin{eqnarray}  \label{omega}
\displaystyle \Omega^4 + \Omega^3 \rho\frac{m+2}{m} -2\Omega^2
\rho^2 \frac{m-1}{m} -4\Omega \frac{\rho^3}{m}-
\nonumber \\
\mathrm{i}ku\frac{\rho}{m^2 r}\left(\Omega^2 \frac{m-1}{2} +\Omega
\frac{\rho m}{2}+\rho^2 m \right)-
\nonumber \\
4\epsilon k^2 \left(\Omega^2  +\Omega \frac{2\rho }{m}+
\mathrm{i}ku\frac{\rho (m-1)}{2m^2 r}\right)=0.
\end{eqnarray}

Below we perform the analysis of this equation in two
qualitatively different limiting cases.

\subsection{Linear viscosity, $m=1$}

The vortex motion in the case $m=1$ can be described in terms of
the magnetic flux viscose flow with the constant anisotropic
viscosity coefficient. Indeed, the current-voltage
characteristics~(\ref{5}), (\ref{mfd}) leads to the linear tensor
relationship between the vortex velocity and the Lorentz driving
force,
\begin{equation}  \label{me}
\Gamma_{ik}N_\alpha V_{\alpha k}=F_{L i}, \qquad {\vec F}_L =
\frac{1}{c}{\vec B}\times {\vec J},
\end{equation}
where the viscosity tensor $\Gamma_{ik}$ is characterized by the
principal values
\begin{equation}\label{pv}
\Gamma_{XX}=\frac{\Phi_0^2}{c^2}\frac{A}{E_0}, \qquad
\Gamma_{YY}=\frac{\Phi_0^2}{c^2 \varepsilon}\frac{A}{E_0}.
\end{equation}
At $m=1$, Eq.~(\ref{omega}) takes on the form,
\begin{eqnarray}\label{1m1}
\Omega^4 + 6r\Omega^3 -32r^3\Omega - 2\mathrm{i}ku r(\Omega +4r)-
\nonumber \\
4\epsilon k^2 \Omega (\Omega +4r)=0.
\end{eqnarray}
It has a simple solution with the following asymptotics at $k
\rightarrow \infty$, $\varepsilon \rightarrow 0$,
\begin{equation}\label{2m1}
\Omega \approx (2k|u|r)^{1/3} \exp(-\mathrm{i}\pi/6).
\end{equation}
The corresponding increment $\lambda$ increases unrestrictedly at
$k \rightarrow \infty$,
\begin{equation}\label{3m1}
\lambda \approx  (k|u|r/4)^{2/3}-\mathrm{i}ku.
\end{equation}
Considering the finite value of $\varepsilon $ limits this
increase to a certain maximum value of Re$\lambda _m$~\cite{prl},
\begin{equation}\label{4m1}
\mathrm{Re}\, \lambda _m \approx \frac{|u|r}{8\varepsilon
^{1/2}}-4r^2, \qquad k_m \approx
\frac{(|u|r)^{1/2}}{(4\varepsilon)^{3/4}}.
\end{equation}

The existence of the solution with $\mathrm{Re}\, \lambda>0$ means
that the instability of the base profiles near the flux-antiflux
interface exists. The increment of the instability grows with an
increase of the wave number if $k<k_m$. Therefore, the
perturbations with the characteristic spatial scale $l_c \sim
1/k_m$ along the $y$-axis is predominant over others. Thus, the
flux-antiflux interface is disturbed in a turbulent-like
manner~\cite{LL} and we can suppose that the presented mechanism
of the macroturbulence could explain the experiment.

As was shown in Ref.~\onlinecite{prl}, the instability exists for
very small $\varepsilon$ only,
\begin{equation}\label{5m1}
\varepsilon \le 0.019 \left[\frac{U} {2RN_0}\right] ^{2}\ll 1.
\end{equation}

So stringent a requirement imposed on the parameter of anisotropy
prompted us to consider the more general situation of arbitrary
$m$. The case $m\gg 1$ is discussed in the next subsection.

\subsection{Power-law current-voltage
characteristics, $m\gg 1$}

The parameter $\epsilon =\varepsilon ^m$ in Eq.~(\ref{omega})
becomes negligible at $m \gg 1$. Therefore, Eq.~(\ref{omega})
transforms into
\begin{eqnarray}\label{1ml}
X^4 + \frac{m+2}{m} X^3 -2\frac{m-1}{m}X^2 -\frac{4}{m}X +
\nonumber \\ {\rm i} \kappa\left(\frac{m-1}{m}X^2 +X +2 \right)=0,
\\
X=\frac{\Omega}{\rho}, \qquad \kappa = k|u|/2mr\rho. \nonumber
\end{eqnarray}
Contrary to Eq.~(\ref{1m1}), this equation contains the term ${\rm
i} \kappa X^2(m-1)/m$ which plays an essential role at $\kappa \gg
1 $ and changes radically the character of the solution:
\begin{equation}\label{2ml}
X \approx \kappa^{1/2}\exp\left(- \frac{{\rm i}
\pi}{4}\right)-\frac{1}{2m}-\frac{2}{\kappa^{1/2}}\exp\left(\frac{{\rm
i}\pi}{4}\right).
\end{equation}
The correspondent increment $\lambda$,
\begin{equation}\label{3ml}
\lambda \approx mr\rho \left({\rm i} \kappa
+2-\frac{\kappa^{1/2}}{2^{1/2}m} \right), \quad m, \kappa \gg 1,
\end{equation}
is almost an imaginary quantity but its relatively small real part
is positive at $\kappa <8 m^2$. This means that the instability
develops as an oscillating process with a magnitude increasing in
time. The wave number $\kappa $ of the instability is bounded from
above due to the finiteness of the parameter $1/m$. Contrary to
the case $m=1$, the instability occurs regardless of the strength
of the current anisotropy, practically, if the anisotropy
parameter $\varepsilon \ll 1$.

\section{Numerical analysis and discussion}

On the basis of the developed theory, let us analyze the
instability of the vortex-antivortex system. In general case of
arbitrary $\varepsilon $ and $m$ this analysis requires that the
dispersion equation (\ref{omega}) be numerically solved.

\begin{figure}[tbp]
\includegraphics[width=0.3\textwidth,height=0.3\textwidth]{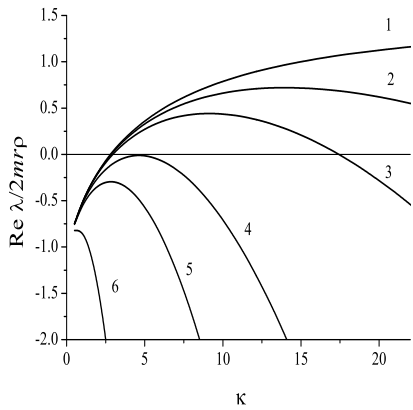}
\caption{\label{f3} The numerical solution $\mathrm{Re}\lambda
(\kappa)$ of the dispersion equation  for $m=20$, $(u/r)^2=0.02$
at different values of the anisotropy parameters $\varepsilon$:
$\varepsilon = 0.5$ (curve 1), $\varepsilon = 0.45$ (2),
$\varepsilon = \varepsilon_c = 0.43$ (3), $\varepsilon  = 0.4$
(4), $\varepsilon = 0.38$ (5), and $\varepsilon = 0.2$ (6).}
\end{figure}

\begin{figure}[!h]
\includegraphics[width=0.3\textwidth,height=0.3\textwidth]{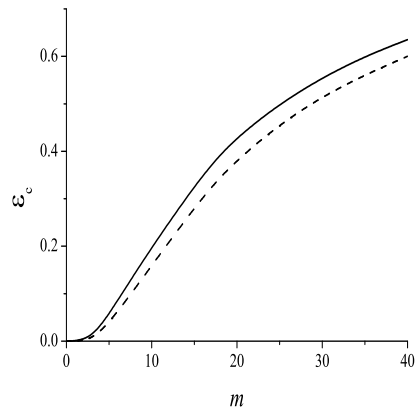}
\caption{\label{f4} The separatrix dividing the stability and
instability regions in the phase plane ($\varepsilon, m$). The
solid line corresponds to $(u/r)^2=0.02$, the dashed line is
obtained for $(u/r)^2=0.002$.}
\end{figure}

The dependence of the increment Re$\lambda $ on the wave number
$\kappa $ for $m=20$ and the different values of the anisotropy
parameter $\varepsilon $ is shown in Fig.~\ref{f3}. For
definiteness, the value of the ratio $(u/r)^2$ is taken to be
equal to 0.02. Remind that this ratio should be small in
accordance with the assumptions made above. The figure shows that
the spectrum of perturbations strongly depends on the parameter of
anisotropy.
Instability is observed at $\varepsilon \leq \varepsilon _c
\approx 0.43$.  Indeed, the increment Re$\lambda $ becomes
positive at some interval of the wave number $\kappa $ and reaches
the maximum value at finite $\kappa $. If the anisotropy parameter
is higher than $\varepsilon _c$, the increment Re$\lambda $ is
negative at any value of $\kappa $. This implies that the
vortex-antivortex system is stable with respect to small
perturbations if the current anisotropy of a superconductor is not
small enough. Of course, the critical value $\varepsilon _c$ of
the anisotropy parameter depends strongly on the exponent $m$. The
graphs in Fig.~\ref{f4} illustrate the change in the value of
$\varepsilon _c$ with an increase of $m$. The function
$\varepsilon _c (m)$ is not very sensitive to the value of the
small parameter $u/r$ in our theory. The graph $\varepsilon _c
(m)$ represents the separatrix dividing the phase space
$(m,\varepsilon)$ into two parts corresponding to the stable
(left-hand part) and unstable (right-hand part) states of the
vortex system.

\begin{figure}[!h]
\includegraphics[width=0.4\textwidth,height=0.4\textwidth]{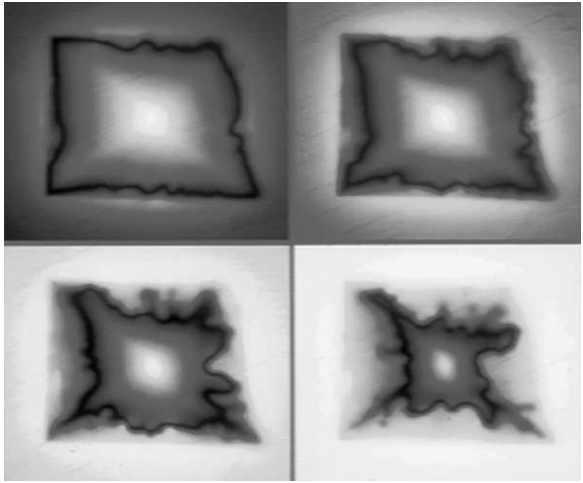}
\caption{\label{f5} The MO image of the development of the
macroturbulence in a detwinned YBCO single crystal.}
\end{figure}

One of the most important results of this paper consists in the
substantial weakening the requirement theoretically imposed on the
anisotropy parameter $\varepsilon$ for the observation of the
instability in the vortex-antivortex system. The effective
parameter of the anisotropy determining the existence of the
instability is $\varepsilon ^m$ in the considered case of the
nonlinear current-voltage characteristics. Since the exponent $m$
can reach several tens in real superconductors, the necessary
condition for the instability is achieved for not too small values
of $\varepsilon$.
This circumstance allows one to eliminate a seeming contradiction
between our theory and experiments where the macroturbulence was
reported to be observed even in detwinned YBCO single crystals.
The MO image of the development of the macroturbulence in such a
sample is shown in Fig.~\ref{f5}. The twinning structure in this
sample is not observed in polarized light whereas the
macroturbulence is clearly pronounced. However, a careful scan of
the left top image in Fig.~\ref{f5}, where the initial magnetic
flux distribution is shown, highlights the presence of some
anisotropy of the current-carrying capability in the sample plain.
The penetration depth in the horizontal direction is clearly seen
to be about 1.2 times higher than in the vertical one. This means
that the anisotropy exists and the macroturbulent instability may
appear. Such a situation with the detwinned samples is, perhaps,
typical for the 1--2--3 single crystals. After the detwinning
procedure, some traces of the twin structure remain inside a
sample and the anisotropic distribution of the impurities takes
place. The existence of the anisotropy of the electrical
resistivity in the \textbf{ab} plane of detwinned YBCO single
crystals was reported in Ref.~\onlinecite{kwok}. Irrespective of
its nature, the anisotropy in the detwinned samples can cause the
observed macroturbulent instability.

Finally, we should note that the qualitative features of the
studied instability are quite similar at any $m>1$.

\section{Conclusion}
The very interesting phenomenon of the macroturbulent instability
has so far been observed only in superconductors of 1--2--3
systems. Such systems are normally characterized by a
well-pronounced anisotropy of the current-carrying capability in
the \textbf{ab} plane. This experimental fact provided the basis
for the theoretical approach to explain the nature of the
instability. Under the action of the Lorentz force, the vortices
and antivortices move towards each other at some non-zero angle
with respect to the front of magnetization reversal owing to the
anisotropy. As a result, the vortex and antivortex hydrodynamic
flow is characterized by the tangential discontinuity near the
front. Specifically this fact leads to the turbulization. The
anisotropy is described in terms of the power-law anisotropic
current-voltage characteristics. The analysis of the linearizes
set of equations consisting of the continuity and Maxwell
equations along with appropriate boundary conditions allowed us to
derive the dispersion equation for the increment of the
instability. As is shown, the instability exists in a wide range
of the parameters of the problem. In particular, the instability
can be observed at not very strong anisotropy if the
current-voltage characteristics is steep. This conclusion agrees
with the experiment.

Unfortunately, we cannot make a direct comparison between the
theoretical results and the experimental data because the theory
operates with the unknown phenomenological parameter $R$
describing the vortex-antivortex annihilation rate. In order to
express this parameter in terms of observable variables, the
microscopic model of the annihilation process needs to be
constructed. Nevertheless, the proposed theory and the obtained
phase diagram, showing the region where the instability occurs,
can be useful for the qualitative analysis of the macroturbulence.
For instance, the existence of the instability in a definite
temperature region alone can be simply rationalized within our
model. At low temperatures, the critical current density
increases, and the characteristic spatial scale $L$, in
Eq.~(\ref{dv}), decreases correspondingly and becomes comparable
to or less than the twin-boundary spacing. As a result, the
anisotropy is suppressed and the instability disappears. On the
other hand, at temperatures close to $T_{c}$ the anisotropy is no
longer effective due to the thermal activation of the vortices. It
is a remarkable thing fully supporting our model, that in the
present heavily twinned crystal the turbulence occurs at much
lower temperatures than is found in previous studies of similar
crystals with only little twinning~\cite{koblisch,joh}.

\begin{acknowledgments}
This work is supported by INTAS (grant 02--2282), RFBR (grants
00-02-17145, 00-02-18032), Russian National Programm on
Superconductivity (contract~40.012.1.1.11.46), and the Research
Council of Norway.
\end{acknowledgments}

\end{document}